\documentclass[twocolumn,english,aps,pra,showpacs,superscriptaddress]{revtex4}
\usepackage{pslatex}
\usepackage[T1]{fontenc}
\usepackage[latin1]{inputenc}
\usepackage{amssymb}

\makeatletter

\providecommand{\LyX}{L\kern-.1667em\lower.25em\hbox{Y}\kern-.125emX\@}

\usepackage{babel}
\usepackage{graphicx}
\makeatother
\input{epsf}
\begin{document}

\title{Non-divergent pseudo-potential treatment of 
spin-polarized fermions under 1D and 3D harmonic
confinement}

\author{K. Kanjilal and D. Blume}
\affiliation{Department of Physics, Washington State
University, Pullman, WA 99164-2814}

\begin{abstract}
Atom-atom scattering of bosonic one-dimensional (1D) atoms 
has been modeled successfully using a zero-range delta-function potential, while
that of bosonic 3D atoms has been
modeled successfully 
using Fermi-Huang's regularized $s$-wave pseudo-potential. 
Here, we derive the eigenenergies of
two spin-polarized 1D fermions 
under external harmonic confinement interacting through a zero-range
potential, which only acts on odd-parity wave functions, analytically.
We also present a  divergent-free zero-range potential treatment
of two spin-polarized 3D fermions under harmonic confinement.
Our pseudo-potential treatments are verified through
numerical calculations for
short-range model potentials.
\end{abstract}

\pacs{34.50.-s,34.10.+x}         

\maketitle

\section{Introduction}
Recently,
atom-atom scattering has received renewed interest since
the properties of ultracold atomic (bosonic or fermionic) gases
depend predominantly on a single atom-atom scattering parameter~\cite{dalf98}. 
This is 
the $s$-wave
scattering length $a_s$ for a three-dimensional (3D) Bose gas~\cite{ande95} (or
for a 3D Fermi gas consisting of atoms with ``spin-up'' and ``spin-down''), 
and the 
$p$-wave scattering volume $V_p$~\cite{rega03,suno03} for a 
3D spin-polarized Fermi gas.
For a 1D or quasi-1D gas, it is the 1D scattering 
length $a_{1D}$~\cite{mattis,olsh98}, 
which characterizes
the
even-parity and odd-parity 
spatial wave function applicable to 
bosons and to spin-polarized fermions, respectively.
In many instances, atom-atom scattering processes
can be conveniently modeled through
a shape-independent pseudo-potential~\cite{demk88,albe88}, 
whose coupling strength 
is chosen such that it reproduces 
the scattering properties of the full shape-dependent
3D or 1D atom-atom potential.

Fermi-Huang's regularized 
pseudo-potential~\cite{ferm34,brei47,huan57} 
supports a single bound state
for positive $a_s$ and no bound state for negative $a_s$. It
has been used frequently to describe 3D $s$-wave scattering
between two bosons or two fermions with different generalized spin. 
Busch {\em{et al.}}~\cite{busc98}, e.g.,
derive the eigenenergies 
for
two atoms under harmonic confinement 
interacting through
Fermi Huang's pseudo-potential
analytically.
Using an energy-dependent 
scattering length $a_s(E)$, their results
can be applied successfully to situations
where $a_s$ is large and
positive, i.e., near
a Feshbach resonance~\cite{ties00,blum02,bold02}.
Building on these results, Borca~{\em{et al.}}~\cite{borc03} 
use a simple two-atom model 
to explain many aspects of an experiment that produces molecules
from a sea of cold atoms using magnetic field ramps~\cite{donl02}.
In addition to these two-body applications,
Fermi-Huang's 3D $s$-wave 
pseudo-potential 
plays a key role in developing (effective) many-body theories.

This paper determines the eigenspectrum of 
two spin-polarized 3D fermions interacting through a regularized 
$p$-wave zero-range potential, parameterized through a {\em{single
parameter}}, i.e., the 
$p$-wave scattering volume $V_p$, under
harmonic confinement analytically.
Since wave functions with relative angular momentum $l$
greater than zero
have vanishing amplitude at $r=0$ (where $r$ denotes the distance between
the two atoms), our zero-range 
$p$-wave potential contains
derivative operators.
Furthermore, it
contains, following ideas suggested by Huang and Yang in 1957~\cite{huan57}, 
a so-called regularization
operator, which eliminates divergencies at $r=0$ that would arise otherwise.
We show that our pseudo-potential
imposes
a boundary condition on
the wave function at $r=0$ (see also Ref.~\cite{huan89});  
this boundary condition serves as an alternative representation
of the $p$-wave pseudo-potential. 
Earlier studies, in contrast, impose a boundary
condition at finite $r$, corresponding to a 
{\em{finite-range}} pseudo-potential
with {\em{two parameters}}~\cite{demk81,frol03}.
The validity of 
our
pseudo-potential
is demonstrated by comparing the eigenenergies determined analytically
for two particles under harmonic confinement with those
determined numerically
for shape-dependent atom-atom potentials.

Due to significant 
advancements in trapping and cooling, to date
cold atomic gases cannot only be trapped in 3D geometries but also 
in quasi-2D and quasi-1D geometries~\cite{reic99,muel99,goer01}. 
In the quasi-1D regime, the
transverse motion is ``frozen out'' so that the behaviors of atomic gases
are dominated by the longitudinal motion.
Quasi-1D gases can hence often be
treated within a 1D model, where the atoms are restricted
to a line. To model 1D atom-atom interactions,
for which the spatial wave function has {\em{even parity}}, 
delta-function contact interactions have been used successfully. 
In contrast to the 3D $s$-wave delta-function potential, which requires
a regularization, the 1D delta-function
pseudo-potential is non-divergent~\cite{wodk91}. 
To treat spin-polarized 1D fermions, a pseudo-potential that acts 
on spatial wave functions with 
{\em{odd parity}} is needed. Here, we use such a 
pseudo-potential to determine the eigenenergies of two 
spin-polarized 1D fermions 
under harmonic confinement ana\-lytically.
Comparison with eigenenergies determined numerically for shape-dependent
1D atom-atom potentials illustrates the applicability of our 
1D pseudo-potential. 
Our results confirm the Fermi-Bose 
duality~\cite{cheo98,cheo99,gran03,gira03,gros04} 
in 1D for two atoms under harmonic confinement.

\section{Two interacting 1D particles under harmonic confinement}
Consider two 1D atoms with mass $m$ and coordinates
${z}_1$ and ${z}_2$, respectively, under 
external harmonic confinement,
\begin{eqnarray}
V_{trap}({z}_1,{z}_2)= \frac{1}{2}m \omega_z^2 
({z}_1^2+{z}_2^2),
\end{eqnarray}
where $\omega_z$ denotes the angular frequency. 
After separating the center of mass and the relative motion,
the Schr\"odinger equation for the relative degree of freedom ${z}$, 
where ${z}={z}_2-{z}_1$, reads
\begin{eqnarray}
H_{1D} \psi_{1D}({z})  = {E}_{1D} \psi_{1D}({z}),
\label{eq_sebar1D}
\end{eqnarray}
where
\begin{eqnarray}
H_{1D} = -\frac{\hbar^2}{2\mu} \frac{d^2}{d{z}^2} + 
V({z}) + \frac{1}{2}\mu \omega_z^2 {z}^2.
\label{eq_hambar1D}
\end{eqnarray}
Here, $V({z})$ denotes the 1D atom-atom interaction potential,
and $\mu$ the reduced mass, $\mu = m/2$.

Section~\ref{sec1D0} reviews the pseudo-potential 
treatment of two 1D particles with even-parity eigenstates, i.e., 
two bosons or two
fermions with opposite spin, under harmonic confinement.
Section~\ref{sec1Da}
determines the relative eigenenergies
${E}_{1D}^-$ for  
two spin-polarized 1D fermions
interacting through a momentum-dependent zero-range
potential
under harmonic confinement analytically. 
Section~\ref{sec1Db}
benchmarks our treatment of the momentum-dependent zero-range
potential by
comparing with numerical results obtained
for a short-range model potential.

\subsection{Review of pseudo-potential treatment: Even parity}
\label{sec1D0}
The
relative eigenenergies
${E}_{1D}^+$ corresponding to states 
with even parity (in the following referred to as even-parity eigenenergies) 
of two 1D particles 
interacting through the zero-range
pseudo-potential $V_{pseudo}^+({z})$, 
where
\begin{eqnarray}
V_{pseudo}^+({z}) = \hbar \omega_z g_{1D}^+ \delta^{(1)}({z}),
\label{eq_pseudo1Deven}
\end{eqnarray}
have been determined by Busch {\em{et al.}}~\cite{busc98}:
\begin{eqnarray}
\frac{g_{1D}^+}{a_{z}} = -  
\frac{2\Gamma (-\frac{{E}_{1D}^+}{2 \hbar \omega_z} +  \frac{3}{4})  }
{\Gamma      (-\frac{{E}_{1D}^+}{2 \hbar \omega_z} +  \frac{1}{4})}.
\label{eq_1Deven}
\end{eqnarray}
In Eq.~(\ref{eq_pseudo1Deven}), 
$\delta^{(1)}({z})$ denotes the usual 1D delta function.
The transcendental equation~(\ref{eq_1Deven})
allows the 
coupling strength $g_{1D}^+$
for a given energy ${E}_{1D}^+$ to be determined readily.
Vice versa, for a given $g_{1D}^+$, the
even-parity eigenenergies
${E}_{1D}^+$ can be determined semi-analytically.
Figure~\ref{fig1}(a) shows the resulting 
eigenenergies ${E}_{1D}^+$ of two 1D bosons 
or two 1D fermions with opposite spin as a function of the 
coupling strength $g_{1D}^+$.
As expected, for vanishing interaction strength ($g_{1D}^+=0$), 
the relative energies ${E}_{1D}^+$ coincide with
the
harmonic oscillator eigenenergies ${E}_{n}^{osc}$ with even parity,
${E}_{n}^{osc}=(2n+\frac{1}{2})\hbar \omega_z$,
where $n=0,1,\cdots$.

For $|E_{1D}^+| \rightarrow \infty$ (and correspondingly
negative $g_{1D}^+$), Eq.~(\ref{eq_1Deven})
reduces to lowest order to 
\begin{eqnarray}
E_{1D}^+ = -\frac{\hbar^2}{2 \mu (a_{1D}^+)^2},
\label{eq_1Dwo}
\end{eqnarray}
which coincides with
the exact binding energy of the pseudo-potential $V_{pseudo}^+(z)$
without confining potential.
In Eq.~(\ref{eq_1Dwo}),
$a_{1D}^+$ denotes 
the 1D even-parity scattering length, 
\begin{eqnarray}
a_{1D}^+ = \lim_{k \rightarrow 0} -\frac{\tan ( \delta_{1D}^+(k))}{k},
\label{eq_1dscatt}
\end{eqnarray}
which is related to
the 1D coupling constant $g_{1D}^+$ through
\begin{eqnarray}
a_{1D}^+ = - \frac{1}{g_{1D}^+}.
\end{eqnarray}
In Eq.~(\ref{eq_1dscatt}), $k$ denotes the relative 1D wave vector,
$k=\sqrt{2 \mu E_{sc}} / \hbar$, and $E_{sc}$ the 1D scattering energy.
The phase shift $\delta_{1D}^+$ is obtained by matching the 
free-space scattering solution for positive $z$ to 
$\sin( k z + \delta_{1D}^+)$. 
The dashed line in Fig.~\ref{fig1}(a) shows
the binding energy of the even-parity pseudo-potential without
confinement, Eq.~(\ref{eq_1Dwo}), while the dash-dotted line shows the 
expansion of Eq.~(\ref{eq_1Deven}) to next higher order.

In addition to the
1D eigenenergies $E_{1D}^+$, the eigen functions 
$\psi_{1D}^+(z)$ can be determined analytically, resulting in the
logarithmic
derivative
\begin{eqnarray}
\left[
\frac{ \frac{d \psi_{1D}^+(z)}{dz}}{\psi_{1D}^+(z)}
\right]_{z \rightarrow 0^+}
= 
\frac{g_{1D}^+}{a_z^2}.
\label{eq_bc1Deven}
\end{eqnarray}
This boundary condition 
is an alternative representation of the
even-parity
pseudo-potential $V_{pseudo}^+(z)$.

\subsection{Analytical pseudo-potential treatment: Odd parity}
\label{sec1Da}
Following the derivation of the even-parity eigenenergies by 
Busch {\em{et al.}}~\cite{busc98}, 
we now derive an analogous expression for the odd-parity
eigenenergies ${E}^-_{1D}$ using the
zero-range pseudo-potential $V_{pseudo}^-({z})$, 
\begin{eqnarray}
V_{pseudo}^-({z}) = 
\hbar \omega_z {g_{1D}^-} 
{\frac{ ^{\leftarrow}d }        { d{z}} }
\delta^{(1)}({z}) 
\frac{ d^{\rightarrow}} {d{z}}.
\label{eq_pseudoodd}
\end{eqnarray}
This pseudo-potential leads to
discontinuous eigenfunctions with continuous derivatives at ${z}=0$.
We show that the 
logarithmic derivative
of $\psi_{1D}^-(z)$ is well-behaved 
for $z \rightarrow 0^+$.
In Eq.~(\ref{eq_pseudoodd}), 
the first derivative acts to the left and the second 
to the right, 
\begin{eqnarray}
\int_{-\infty}^{\infty} \phi^*({z})  
V_{pseudo}^-({z})  \chi({z}) d{z} = 
\hbar \omega_z g_{1D}^-
\frac{d \phi^*(0)}{d {z}} \frac{d \chi(0)}{d {z}},
\end{eqnarray}
with the short-hand notation 
\begin{eqnarray}
\frac{d\chi(0)}{d{z}}
= \left[ \frac{d \chi({z})}{d{z}} \right]_{z=0}.
\end{eqnarray}
Since $V_{pseudo}^-({z})$
acts only on wave functions with odd parity
(and not on those with even parity),
we refer 
to $V_{pseudo}^-({z})$ 
as odd-parity pseudo-potential;
however, $V_{pseudo}^-({z})$ itself has even
parity.
Similar pseudo-potentials have
recently also been used by others~\cite{gira03,gros04,gira04}.

To start with, we expand the {\em{discontinuous}}
odd-parity eigenfunction $\psi_{1D}^-(z)$ 
in {\em{continuous}}
1D odd-parity harmonic oscillator eigenfunctions $\phi_n(z)$,
\begin{eqnarray}
\psi_{1D}^-(z)  = \sum_{n=0}^{\infty} c_n  \phi_n(z) ,
\label{eq_expand1D}
\end{eqnarray}
where the $c_n$ denote expansion coefficients, and 
\begin{eqnarray}
\phi_n(z) = 
\sqrt{
\frac{{2}}{{L_{n}^{(1/2)}(0)} {\sqrt{\pi} \; a_z}
}} \;\;
\frac{z}{a_z} \; 
\exp \left(-\frac{z^2}{2 a_z^2} \right) \, L_{n}^{(1/2)}\left(
\frac{z^2}{a_z^2} \right),
\label{eq_1Dho}
\end{eqnarray}
where $a_{z}=\sqrt{\hbar/(\mu \omega_z)}$.
In Eq.~(\ref{eq_1Dho}), the $L_n^{(1/2)}\left(z^2/a_z^2 \right)$ 
denote associated Laguerre polynomials and
the $\phi_n(z)$ are normalized to one,
\begin{eqnarray}
\int_{-\infty}^{\infty} |\phi_n(z)|^2 dz =1.
\end{eqnarray}
The corresponding odd-parity harmonic oscillator eigenenergies are
\begin{eqnarray}
E_{n}^{osc}=\left(2n+\frac{3}{2} \right) \hbar \omega_z,
\label{eq_1Dhoodd}
\end{eqnarray}
where $n=0,1,\cdots$.
Inserting expansion (\ref{eq_expand1D})
into Eq.~(\ref{eq_sebar1D}), multiplying from the left with $ \phi_{n'}^*(z)$,
and integrating over $z$,
results in
\begin{eqnarray}
c_{n'} (E_{n'}^{osc} - E_{1D}^-) + \nonumber \\ 
g_{1D}^- \, \hbar \omega_z \, \frac{d \phi^*_{n'}(0)}{dz} \left[
\frac{d}{dz} \left( \sum_{n=0}^{\infty} c_n \phi_n(z) \right)
\right]_{z \rightarrow 0^+} = 0
\label{eq_1Dexco}.
\end{eqnarray}
The coefficients $c_{n'}$ are hence of the form
\begin{eqnarray}
c_{n'} = A \; \frac{\frac{d \phi_{n'}^*(0)}{dz}}{E_{n'}^{osc}-E_{1D}^-},
\end{eqnarray}
where the constant $A$ is independent of $n'$.
Inserting this expression for the $c_{n}$ into Eq.~(\ref{eq_1Dexco}) leads
to
\begin{eqnarray}
\left[ \frac{d}{dz} \left(
\sum_{n =0}^{\infty} 
\frac{\frac{d\phi_n^*(0)}{dz} \phi_n(z)}{E_n^{osc}-E_{1D}^-}
\right) \right]_{z \rightarrow 0^+}
= - \frac{1}{g_{1D}^- \, \hbar \omega_z}. \label{eq_sum1D}
\end{eqnarray}
If we define a non-integer quantum number $\nu$ through
\begin{eqnarray}
E_{1D}^- = \left(2\nu + \frac{3}{2} \right) \hbar \omega_z,
\label{eq_1Dnonint}
\end{eqnarray}
and use expression~(\ref{eq_1Dho}) 
for the $\phi_n(z)$,
Eq.~(\ref{eq_sum1D}) can be rewritten as
\begin{eqnarray}
\frac{1}{\sqrt{\pi}} \left[ \frac{d}{dz} \left\{
z \exp\left(-\frac{z^2}{2 a_z^2} \right)
\sum_{n=0}^{\infty} \frac{L_{n}^{(1/2)}\left( z^2/a_z^2 \right)}{n-\nu}
\right\} \right]_{z \rightarrow 0^+}
= - \frac{a_z^3}{g_{1D}^-}, \label{eq_suma1D}
\end{eqnarray}
where the $z \rightarrow 0^+$ limit 
is well-behaved.
Equation~(\ref{eq_suma1D}) can be
evaluated using the identity
\begin{eqnarray}
\sum_{n=0}^{\infty} \frac{L_{n}^{(1/2)}\left(z^2/a_z^2\right)}{n-\nu} =
\Gamma(-\nu) \; U \left(-\nu,\frac{3}{2},\frac{z^2}{a_z^2} \right),
\label{eq_hyper21D}
\end{eqnarray}
and the known small $z$
behavior of the hypergeometric function
$U\left(-\nu,\frac{3}{2},\frac{z^2}{a_z^2}\right)$~\cite{abraS1},
\begin{eqnarray}
-\frac{1}{\pi} 
U\left(-\nu,\frac{3}{2},\frac{z^2}{a_z^2}\right) \rightarrow \nonumber \\
- 
\frac{1}{\Gamma(-\nu) \Gamma(\frac{1}{2}) } \left(\frac{z}{a_z}
\right)^{-1}  
+\frac{1}{\Gamma(-\nu-\frac{1}{2}) \Gamma(\frac{3}{2})} 
+ {\cal{O}}(z). \label{eq_hyper1D}
\end{eqnarray}
Using Eqs.~(\ref{eq_hyper21D}) and (\ref{eq_hyper1D}) in
Eq.~(\ref{eq_suma1D}),
evaluating the derivative with
respect to $z$, and then taking the $z \rightarrow 0^+$ limit, results in
\begin{eqnarray}
-\frac{a_{z}^3}{g_{1D}^-}= -\frac{\sqrt{\pi}}{\Gamma(\frac{3}{2})} 
\frac{\Gamma(-\nu)}
{\Gamma(-\nu-\frac{1}{2}) } .\label{eq_final1D}
\end{eqnarray} 
Replacing the non-integer quantum number $\nu$ 
[see Eq.~(\ref{eq_1Dnonint})] 
by 
$E_{1D}^-/2 \hbar \omega_z-3/4$, we obtain the transcendental equation
\begin{eqnarray}
\frac{g_{1D}^-}{a_{z}^3}=
\frac{\Gamma(-\frac{E_{1D}^-}{2 \hbar \omega_z}+\frac{1}{4})}
{2 \Gamma(-\frac{E_{1D}^-}{2 \hbar \omega_z}+\frac{3}{4})}, \label{eq_finala1D}
\end{eqnarray}
which allows the 1D odd-parity eigenenergies $E_{1D}^-$
to be determined for a given interaction strength $g_{1D}^-$.

\begin{figure}[tbp]
\vspace*{+.3in}
\centerline{\epsfxsize=3.25in\epsfbox{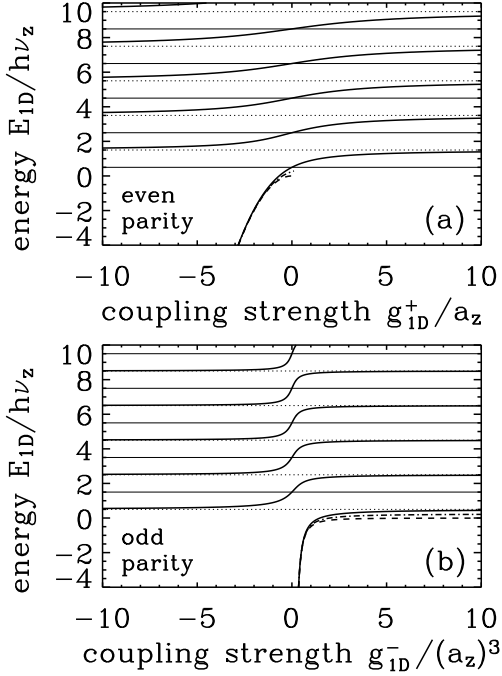}}
\vspace*{-.3in}
\caption{
Solid lines in panel (a) show
the relative even-parity energies $E_{1D}^+$  [Eq.~(\ref{eq_1Deven})]
calculated using the pseudo-potential
$V_{pseudo}^+(z)$
as a function of $g_{1D}^+$.
Solid lines in panel (b) show
the relative odd-parity energies $E_{1D}^-$  [Eq.~(\ref{eq_finala1D})]
calculated using the pseudo-potential
$V_{pseudo}^-(z)$
as a function of $g_{1D}^-$.
Horizontal solid lines indicate the 
harmonic oscillator eigenenergies [with
even parity in panel (a), and
with 
odd parity in panel (b)].
Horizontal dotted lines indicate the asymptotic value of the eigenenergies 
$E_{1D}^+$ 
and $E_{1D}^-$
for $g_{1D}^+ \rightarrow \pm \infty$ and  
$g_{1D}^- \rightarrow \pm \infty$, respectively.
Dashed lines show the binding energies 
$E_{1D}^+$, Eq.~(\ref{eq_1Dwo}), in panel (a) and 
$E_{1D}^-$, Eq.~(\ref{eq_ebindodd}), in panel (b)
of the pseudo-potentials
$V_{pseudo}^+(z)$ and $V_{pseudo}^-(z)$, respectively,
without confinement.
Dash-dotted lines show the expansion of Eq.~(\ref{eq_1Deven}) [panel (a)] and 
Eq.~(\ref{eq_finala1D}) [panel (b)] including the next order term.
}
\label{fig1}
\end{figure}

Solid lines in Fig.~\ref{fig1}(b) 
show the 1D odd-parity eigenenergies $E_{1D}^-$, 
Eq.~(\ref{eq_finala1D}), as a function
of $g_{1D}^-$.
For $g_{1D}^-=0$, the eigenenergies $E_{1D}^-$ coincide with the odd-parity
harmonic oscillator eigenenergies $E_n^{osc}$,
Eq.~(\ref{eq_1Dhoodd}); 
they increase for positive
$g_{1D}^-$ (``repulsive interactions''), and 
decrease for negative $g_{1D}^-$ (``attractive interactions'').

Expansion of Eq.~(\ref{eq_finala1D}) to lowest order
for large and negative eigenenergy (implying positive $g_{1D}^-$),
$|E_{1D}^-| \rightarrow \infty$, results in
\begin{eqnarray}
E_{1D}^- = -\frac{\hbar^2}{2 \mu (a_{1D}^-)^2},
\label{eq_ebindodd}
\end{eqnarray}
where
the 1D scattering length $a_{1D}^-$ is defined analogously to $a_{1D}^+$
[with the superscript ``$+$'' in Eq.~(\ref{eq_1dscatt}) 
replaced by the superscript ``$-$''].
The 1D scattering length $a_{1D}^-$ 
is related to the 1D coupling strength $g_{1D}^-$
through
\begin{eqnarray}
{g_{1D}^-}= a_{1D}^- a_z^2. \label{eq_1Drelation}
\end{eqnarray}
The energy given by Eq.~(\ref{eq_ebindodd}) 
coincides with the binding energy of the 1D
pseudo-potential $V_{pseudo}^-(z)$ 
without
the confining
potential.
A dashed line in Fig.~\ref{fig1}(b) shows $E_{1D}^-$, Eq.~(\ref{eq_ebindodd}),
while a dash-dotted line shows the expansion 
of 
Eq.~(\ref{eq_finala1D}) including the next order term.

In addition to the eigenenergies $E_{1D}^-$,
we calculate the eigenfunctions $\psi_{1D}^-$,
\begin{eqnarray}
\psi_{1D}^-(z) \propto \frac{\Gamma(-\nu)}{\sqrt{a_z}} \,
\frac{z}{a_z} \, \exp \left(-\frac{z^2}{2a_z^2} \right) \, 
U\left(-\nu, \frac{3}{2},\frac{z^2}{a_z^2}\right).
\end{eqnarray}
Following steps similar to those outlined above,
the logarithmic derivative at $z\rightarrow 0^+$ reduces to
\begin{eqnarray}
\left[ \frac{\frac{d \psi_{1D}^-(z)}{dz}}{\psi^-_{1D}(z)}
\right] _{z \rightarrow 0^+} = 
-\frac{a_z^2}{g_{1D}^-}. \label{eq_bc1Dodd}
\end{eqnarray}
Equation~(\ref{eq_bc1Dodd})
is an alternative representation of the 1D odd-parity 
pseudo-potential $V_{pseudo}^-(z)$~\cite{gira03,gros04,gira04}.

The even-parity eigenenergies $E_{1D}^+$ 
[Eq.~(\ref{eq_1Deven})] and the odd-parity eigenenergies 
$E_{1D}^-$ [Eq.~(\ref{eq_finala1D})], as well as the logarithmic derivatives
[Eqs.~(\ref{eq_bc1Deven}) and (\ref{eq_bc1Dodd})]
are identical
if the
coupling constants of $V_{pseudo}^+(z)$ and 
$V_{pseudo}^-(z)$ are chosen as follows,
\begin{eqnarray}
g_{1D}^- =- \frac{a_{z}^4}{g_{1D}^+}.
\end{eqnarray}
This implies that
even-parity energies $E_{1D}^+$ can be
obtained by solving the 1D Schr\"odinger equation, Eq.~(\ref{eq_sebar1D}),
for $H_{1D}$ given by Eq.~(\ref{eq_hambar1D}) with $V(z)=V_{pseudo}^-(z)$ 
[and vice versa,
odd-parity energies $E_{1D}^-$ can be
obtained by solving the 1D Schr\"odinger equation
with $V(z)=V_{pseudo}^+(z)$]. 
Our analytical treatment of two 1D particles under external confinement
thus confirms 
the Fermi-Bose
duality for two 1D particles under harmonic 
confinement~\cite{cheo98,cheo99,gran03,gira03,gros04}.

\subsection{Comparison with shape-dependent 1D atom-atom potential}
\label{sec1Db}
\begin{figure}[tbp]
\vspace*{0.3in}
\centerline{\epsfxsize=3.25in\epsfbox{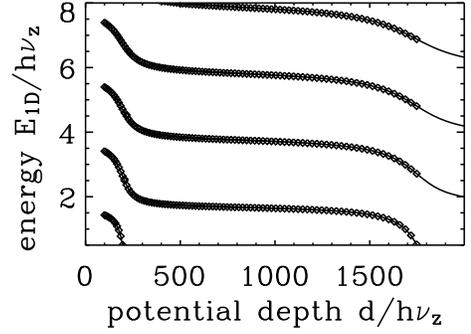}}
\vspace*{-2.1in}
\caption{
Relative odd-parity eigenenergies $E_{1D}^-$  for two
particles under 1D harmonic confinement as a function
of the well depth $d$.
Solid lines show the eigenenergies obtained by solving the
1D Schr\"odinger equation, Eq.~(\ref{eq_sebar1D}), 
for the Hamiltonian given in Eq.~(\ref{eq_hambar1D}) 
numerically
using 
a short-range model potential, Eq.~(\ref{eq_1Dmorse}), for a series
of well depths $d$.
Symbols show the eigenenergies obtained for the pseudo-potential
$V_{pseudo}^-(z)$, 
taking the
energy-dependence of the 1D coupling constant $g_{1D}^-$ into account,
$g_{1D}^-=g_{1D}^-(E_{sc})$ (see text). 
}
\label{fig2}
\end{figure}

To benchmark the applicability of the odd-parity
pseudo-potential  $V_{pseudo}^-(z)$ to two 1D atoms under harmonic
confinement, we solve the 1D Schr\"odinger equation,
Eq.~(\ref{eq_sebar1D}), for the Hamiltonian given by 
Eq.~(\ref{eq_hambar1D}) numerically for the shape-dependent Morse potential
$V_{morse}({z})$,
\begin{eqnarray}
V_{morse}({z}) = d e^{-\alpha({z}-{z}_0)} 
\left[ e^{-\alpha({z}-{z}_0)}-2\right].
\label{eq_1Dmorse}
\end{eqnarray}
Our numerical calculations are performed for a fixed range parameter
${z}_0$, ${z}_0=11.65$a.u., and for $\alpha=0.35$a.u.;
these parameters
roughly approximate the 3D Rb$_2$ triplet potential~\cite{esry99}.
The angular trapping frequency $\omega_{z}$ is fixed at $10^{-9}$a.u.
($2 \pi \, \nu_z = \omega_z$), and the
atom mass $m$ at that of the $^{87}$Rb atom,
implying
an oscillator length $a_z$ of 112.5a.u., and hence
a fairly tightly trapped atom pair.
To investigate potentials with different 1D scattering properties,
we choose depth parameters $d$ 
for which the 1D
Morse potential supports between zero and two 1D odd-parity 
bound states.
Solid lines in Fig.~\ref{fig2} show the resulting
1D odd-parity eigenenergies $E_{1D}^-$
obtained numerically
as a function of $d$. 
The corresponding eigenstates 
have ``gas-like character'', that is, these states would 
correspond to continuum states
if the confining potential was absent.

To compare the odd-parity eigenenergies obtained numerically for the Morse
potential $V_{morse}(z)$
with those obtained for the odd-parity pseudo-potential 
$V_{pseudo}^-(z)$, we 
follow Refs.~\cite{blum02,bold02}.
We first 
perform scattering calculations for the
1D Morse potential (no confinement) as a function of the 
relative scattering energy 
$E_{sc}$
for various depths $d$, which provide, for a given $d$, 
the energy-dependent 1D scattering length $a_{1D}^-(E_{sc})$,
where $a_{1D}^-(E_{sc})= -\tan(\delta_{1D}^-(k))/k$.
Using the relation between the 1D scattering length $a_{1D}^-$ and the
1D coupling strength $g_{1D}^-$, Eq.~(\ref{eq_1Drelation}),
we then solve the transcendental equation~(\ref{eq_finala1D}) self-consistently
for 
$E_{1D}^-$. 

Diamonds in Fig.~\ref{fig2} show the resulting odd-parity 
eigenenergies $E_{1D}^-$ for
two 1D particles under harmonic confinement
interacting through the odd-parity 
energy-dependent pseudo-potential $V_{pseudo}^-(z)$ with 
$g_{1D}^-=g_{1D}^-(E_{sc})$.
Excellent agreement between these eigenenergies
and those 
obtained for the Morse potential (solid lines) 
is visible for all well depths $d$.
We emphasize that this 
agreement depends crucially on the usage of
{\em{energy-dependent}} 1D coupling constants.
In summary, Fig.~\ref{fig2} illustrates that the odd-parity pseudo-potential 
$V_{pseudo}^-(z)$
provides a good description of 
the eigenstates of
two spin-polarized 1D fermions under harmonic confinement
for all interaction strengths, including $g_{1D}^- \rightarrow \pm \infty$.

\section{Two interacting 3D particles under harmonic confinement}
Consider two 3D particles with mass $m$ and 
coordinates $\vec{r}_1$ and
$\vec{r}_2$, respectively, confined by the potential 
$V_{trap}(\vec{r}_1,\vec{r}_2)$,
\begin{eqnarray}
V_{trap}(\vec{r}_1,\vec{r}_2)=\frac{1}{2} \mu \omega_{ho}^2 
\left( \vec{r}_1^2
+\vec{r}_2^2 \right),
\end{eqnarray}
where 
$\omega_{ho}$ denotes the angular trapping frequency of the
harmonic 3D confinement.
The corresponding Schr\"odinger equation
decouples into a center of mass part,
whose solution can be readily
written down, and into a relative part,
\begin{eqnarray}
H_{3D}= H_{3D}^{osc}+
 V(\vec{r}). \label{eq_h3D}
\end{eqnarray}
Here, $\vec{r}$
denotes
the relative coordinate vector ($\vec{r}= \vec{r}_2-\vec{r}_1$), 
$V(\vec{r})$ the atom-atom interaction potential, and
$H_{3D}^{osc}$ the 3D harmonic oscillator Hamiltonian,
\begin{eqnarray}
H_{3D}^{osc}= -\frac{\hbar^2}{2 \mu} \nabla_{\vec{r}}^2
+
\frac{1}{2} \mu \omega_{ho}^2 \vec{r}^2.
\end{eqnarray}
The corresponding Schr\"odinger equation for the relative coordinate reads
\begin{eqnarray}
H_{3D}  \psi_{3D}(\vec{r}) 
= E_{3D}  \psi_{3D}(\vec{r}) .
\label{eq_se3D3D}
\end{eqnarray}

Section~\ref{sec_3d0} briefly reviews Fermi Huang's regularized 
$s$-wave pseudo-potential, 
while Section~\ref{sec3Da} solves Eq.~(\ref{eq_se3D3D})
for a regularized $p$-wave 
zero-range potential analytically.
To illustrate the applicability of this 
$p$-wave pseudo-potential,
Section~\ref{sec3Db} compares the resulting relative eigenenergies $E_{3D}$
for two particles under harmonic confinement with those obtained 
numerically 
for a shape-dependent short-range model potential.

\subsection{Review of 3D pseudo-potential treatment: $s$-wave}
\label{sec_3d0}
Using Fermi-Huang's regularized $s$-wave ($l=0$) pseudo-potential
$V_{pseudo}^{l=0}(\vec{r})$~\cite{ferm34,huan57},
\begin{eqnarray}
V_{pseudo}^{l=0}(\vec{r})= \frac{2 \pi \hbar^2}{\mu} a_s \delta^{(3)}(\vec{r}) 
\frac{\partial}{\partial r} r, \label{eq_pseudoswave}
\end{eqnarray}
where $\delta^{(3)}(\vec{r})$ 
denotes the radial component of the 3D $\delta$-function,
\begin{eqnarray}
\delta^{(3)}(\vec{r}) = \frac{1}{4 \pi r^2} \delta^{(1)}(r),
\end{eqnarray}
and $a_s$ the 3D 
$s$-wave scattering length,
Busch {\em{et al.}}~\cite{busc98} 
derive a transcendental equation for the relative 3D
eigenenergies $E_{3D}$,
\begin{eqnarray}
\frac{a_s}{a_{ho}} = 
\frac{\Gamma(-\frac{E_{3D}}{2 \hbar \omega_{ho}} +\frac{1}{4})}
{2\Gamma(-\frac{E_{3D}}{2 \hbar \omega_{ho}} + 
\frac{3}{4})}.\label{eq_3Dswave}
\end{eqnarray}
Here, $a_{ho}$ denotes the oscillator length,
$a_{ho}=\sqrt{\hbar/(\mu \omega_{ho})}$.
Solid lines in Fig.~\ref{fig3}(a) 
show the $s$-wave energies $E_{3D}$ as a function
of $a_s$.
For large and negative $E_{3D}$ (and hence positive $a_s$), 
an expansion of Eq.~(\ref{eq_3Dswave})
to lowest order results in
\begin{eqnarray}
E_{3D} = -\frac{\hbar^2}{2 \mu (a_s)^2},
\label{eq_3Dbindswave}
\end{eqnarray}
which corresponds to the binding energy of $V_{pseudo}^{l=0}(\vec{r})$
without the confining potential.
A dashed
line in Fig.~\ref{fig1}
shows the energy given by Eq.~(\ref{eq_3Dbindswave}), while a dash-dotted
line shows the expansion of Eq.~(\ref{eq_3Dswave}) including the next higher
order term.

Since only $s$-wave wave functions have a non-vanishing amplitude at 
$r=0$, Fermi-Huang's regularized pseudo-potential 
leads exclusively to $s$-wave scattering
(no other partial waves are scattered).
Equation~(\ref{eq_3Dswave}) hence applies to
two ultracold bosons under external confinement, 
for which higher even partial waves, such as
$d$- or $g$-waves, are negligible.

Recall that the irregular solution with $l=0$ diverges as $r^{-1}$.
The
so-called regularization operator
$\frac{\partial}{\partial r} r$ of
the pseudo-potential
$V_{pseudo}^s(\vec{r})$, Eq.~(\ref{eq_pseudoswave}), cures this divergence.
The
solutions $\psi_{3D}(\vec{r})$ of two particles under external confinement
obey
the boundary condition
\begin{eqnarray}
\left[ 
\frac{\frac{\partial}{\partial r} \left( r \psi_{3D}(\vec{r})
\right) }{r \psi_{3D}(\vec{r}) }
\right] _{r \rightarrow 0} = -\frac{1}{a_s};
\end{eqnarray}
this boundary condition is an alternative representation of 
$V_{pseudo}^{l=0}(\vec{r})$.

\subsection{Analytical 3D pseudo-potential treatment: $p$-wave}
\label{sec3Da}
The importance of angle-dependent $p$-wave
interactions has recently been demonstrated experimentally for
two potassium atoms
in the vicinity of a magnetic field-dependent $p$-wave Feshbach 
resonance~\cite{tick03}.
Here, we use a $p$-wave
pseudo-potential 
to model {\em{isotropic}} atom-atom
interactions;
treatment of anisotropic interactions
is beyond the scope of this paper.

We use the following
$p$-wave pseudo-potential $V_{pseudo}^{l=1}(\vec{r})$,
\begin{eqnarray}
V_{pseudo}^{l=1}(\vec{r}) = 
g_1
{^{\leftarrow}\nabla}_{\vec{r}} 
\delta^{(3)}(\vec{r}) 
{\nabla}_{\vec{r}}^{{\rightarrow}} 
\frac{1}{2}
\frac{\partial^2}{\partial r ^2 } r^2, \label{eq_pseudo3D}
\end{eqnarray}
where the coupling strength $g_1$ ``summarizes'' the scattering properties
of the original 
shape-dependent atom-atom interaction potential~\cite{omon77,hami02},
\begin{eqnarray}
g_1 = \frac{6 \pi \hbar^2}{\mu}  V_p.
\label{eq_couplingg1}
\end{eqnarray}
Here, $V_p$ denotes the $p$-wave scattering volume~\cite{suno03},
\begin{eqnarray}
V_p = \lim_{k \rightarrow 0} -\frac{\tan(\delta_p(k))}{k^3},
\end{eqnarray}
$\delta_p$ the $p$-wave phase shift,
and $k$ the relative 3D collision wave vector.
Similarly to the 1D odd-parity pseudo-potential $V_{pseudo}^-(z)$,
the first gradient ${\nabla}_{\vec{r}}$ 
with respect to the relative vector $\vec{r}$
acts to the left, while the second one acts to the right,
\begin{eqnarray}
\int \phi^*(\vec{r})  V_{pseudo}^{l=1}(\vec{r})  \chi(\vec{r}) d^3\vec{r} = 
\nonumber \\
g_1
\int \left[ {\nabla}_{\vec{r}} \phi^*(\vec{r}) \right]
 \delta^{(3)}(\vec{r}) 
\left[  {\nabla}_{\vec{r}} 
\left\{ \frac{1}{2} \frac{\partial^2}{\partial r ^2} \left( r^2 \chi(\vec{r}) 
\right) \right\} \right] d^3\vec{r}
.
\end{eqnarray}
Just as the $s$-wave pseudo-potential $V_{pseudo}^{l=0}(\vec{r})$ does not
couple to partial waves with $l \ne 0$,
the $p$-wave pseudo-potential $V_{pseudo}^{l=1}(\vec{r})$
does not couple to partial waves with $l \ne 1$~\cite{roth01}.
Pseudo-potentials of the form $g_1 {^{\leftarrow}} {\nabla}_{\vec{r}} 
\delta^{(3)}(\vec{r}) {\nabla}_{\vec{r}} {^{\rightarrow}}$
have been used by a number of researchers 
before~\cite{omon77,roth01,hami02,hami03};
discrepancies regarding the proper value of the coefficient $g_1$,
however, exist (see, e.g., Ref.~\cite{roth01}).
Here, we introduce the
regularization
operator
$\frac{1}{2} \frac{\partial^2}{\partial r ^2} r^2$ 
[Eq.~(\ref{eq_pseudo3D})], which eliminates
divergencies that would arise otherwise from
the irregular $p$-wave solution (which diverges as $r^{-2}$).
A similar regularization operator has been proposed by Huang and Yang 
in 1957~\cite{huan57};
they, however, use it in conjunction with a coupling parameter $g_1$ 
different from that given by
Eq.~(\ref{eq_couplingg1}).
By comparing with numerical results for a shape-dependent model
potential,
we show that the pseudo-potential $V_{pseudo}^{l=1}(\vec{r})$ describes the
scattering behaviors of two spin-aligned 
3D fermions properly (see Sec.~\ref{sec3Db}).

To determine the relative eigenenergies $E_{3D}$ of two
spin-polarized 3D fermions under harmonic confinement analytically, 
we expand the 3D wave function $\psi_{3D}(\vec{r})$
for fixed angular momentum, $l=1$,
in {\em{continuous}} harmonic 
oscillator eigen functions $\phi_{nlm_l}(\vec{r})$,
\begin{eqnarray}
\psi_{3D}(\vec{r})   =
\sum_{nm_l} c_{nm_l}  \phi_{nlm_l}(\vec{r}), \label{eq_expand3D}
\end{eqnarray}
where the $c_{nm_l}$ denote expansion coefficients.
The $\phi_{nlm_l}(\vec{r})$ depend on the 
principal quantum number $n$, the angular momentum quantum  number $l$, and the
projection quantum number $m_l$,
\begin{eqnarray}
H^{osc}_{3D}  \phi_{nlm_l} (\vec{r})  = 
E_{nl}^{osc}  \phi_{n l m_l}(\vec{r}) 
\end{eqnarray}
and
\begin{eqnarray}
E_{nl}^{osc}= \left(2n + l +\frac{3}{2} \right) \hbar \omega_{ho},
\end{eqnarray}
where $n=0,1,\cdots$; $l=0,1,\cdots,n-1$; and $m_l=0,\pm 1,\cdots,\pm l$.
The $\phi_{nlm_l}(\vec{r})$
can be written
in spherical coordinates [$\vec{r}=(r,\vartheta,\varphi)$],
\begin{eqnarray}
\phi_{nlm_l}(\vec{r})= \sqrt{4 \pi} 
\; R_{nl}(r)  \;
Y_{l m_l}(\vartheta,\varphi),
\end{eqnarray}
where the $Y_{lm_l}(\vartheta,\varphi)$ denote spherical harmonics and
the $R_{nl}(r)$ are given by
\begin{eqnarray}
R_{nl}(r)=
\sqrt{
\frac{2^l}
{(2l+1)!! \; \sqrt{\pi^3} \; L_n^{(l+1/2)}(0) a_{ho}^3} 
}\times \nonumber \\
\left(\frac{r}{a_{ho}}\right)^l \; 
\exp\left(-\frac{r^2}{2 a_{ho}^2} \right) \, 
L_n^{(l+1/2)}\left(\frac{r^2}{a_{ho}^2}\right), \label{eq_rnl}
\end{eqnarray}
with
\begin{eqnarray}
(2l+1)!! = 1 \cdot 3 \cdot \dots \cdot (2l+1).
\end{eqnarray}
The
normalizations of $R_{nl}(r)$ and $Y_{lm_l}(\vartheta,\varphi)$
are chosen as 
\begin{eqnarray}
\int_0^{2\pi} \int_0^{\pi}  |Y_{l m_l}(\vartheta, \varphi)|^2 \;
\sin \vartheta d \vartheta d \varphi = 1
\end{eqnarray}
and
\begin{eqnarray}
\int_0^{\infty} |R_{nl}(r)|^2 \; r^2 dr = \frac{1}{4 \pi}.
\end{eqnarray}

If we plug expansion~(\ref{eq_expand3D}) 
into the 3D Schr\"odinger equation, Eq.~(\ref{eq_se3D3D}),
for the Hamiltonian given by Eq.~(\ref{eq_h3D}) with $V(\vec{r})=
V_{pseudo}^{l=1}(\vec{r})$,
multiply from the left with
$ \phi_{n'lm_l'}^*(\vec{r}) $ [with $l=1$],
and integrate over $\vec{r}$,
we obtain an expression for the coefficients $c_{n'm_l'}$,
\begin{eqnarray}
 c_{n'm_l'} (E_{n'l}^{osc} - E_{3D}) = \nonumber \\
-
g_1
 \left[ {\nabla}_{\vec{r}} R_{n'l}^*(0)
\right] \cdot
 \left[ 
{\nabla}_{\vec{r}} \left\{
\frac{1}{2}\frac{\partial^2}{\partial r^2} \left( 
r^2 \sum_{n=0}^{\infty}
c_{nm_l'}  R_{nl}({r}) \right) \right\} \right]_{r \rightarrow 0} ,
\label{eq_3D42}
\end{eqnarray}
where 
\begin{eqnarray}
{\nabla}_{\vec{r}} R_{nl}^*({0})
=
\left[ {\nabla}_{\vec{r}} R_{nl}^*({r}) \right]
_{{r} = 0}. \label{eq_3Dstep1}
\end{eqnarray}
In deriving Eq.~(\ref{eq_3D42}), we use that
\begin{eqnarray}
\nabla_{\vec{r}}\left[R_{nl}(r) Y_{l m_l}(\vartheta,\varphi)\right]
= \nonumber \\ 
\left[ \nabla_{\vec{r}} R_{nl}(r)\right] Y_{l m_l}(\vartheta,\varphi) +
R_{nl}(r) \left[ \nabla_{\vec{r}}
Y_{l m_l}(\vartheta,\varphi)\right], \label{eq_3Dhelp}
\end{eqnarray}
where the second term on the right-hand side goes 
to zero in the $r \rightarrow 0$ limit.
Since the gradients ${\nabla}_{\vec{r}}$ in Eq.~(\ref{eq_3D42}) 
act on arguments that depend solely on $r$,
we can replace them by
$\hat{e}_{r} \frac{\partial }{\partial r}$
(where $\hat{e}_r$ denotes the unit vector in the $r$-direction),
\begin{eqnarray}
 c_{n'm_l'} (E_{n'l}^{osc} - E_{3D}) = \nonumber \\
-
g_1
 \frac{\partial R_{n'l}^*(0)}{\partial r}
\left[
\frac{1}{2}
\frac{\partial^3}{\partial r^3} \left(r^2 \sum_{n=0}^{\infty}
c_{nm_l'}  R_{nl}({r}) \right) \right]_{r \rightarrow 0} .
\label{eq_3D44}
\end{eqnarray}
Equation~(\ref{eq_3D44}) implies that the coefficients $c_{n' m_l'}$ are
of the form
\begin{eqnarray}
c_{n'm_l'}= {A} \;
\frac{ \frac{\partial R_{n'l}^*(0)}{\partial r}  }
{E_{n'l}^{osc}-E_{3D}},
\label{eq_coeff3d}
\end{eqnarray}
where ${A}$ is a constant 
independent of $n'$.
Plugging Eq.~(\ref{eq_coeff3d})
into
Eq.~(\ref{eq_3D44}) results in 
an implicit expression for the 3D energies $E_{3D}$,
\begin{eqnarray}
\left[\frac{1}{2}
 \frac{\partial^3}{\partial r^3} \left( r^2
\sum_{n=0}^{\infty} 
\frac{  \frac{\partial R_{nl}^*({0})}{\partial r}
R_{nl}(r)}
{E_{nl}^{osc}-E_{3D}} \right)
\right]_{r \rightarrow 0} = 
-
\frac{1}{g_1}.
\label{eq_sum3D}
\end{eqnarray}
To simplify the infinite sum over $n$,
we use expression~(\ref{eq_rnl}) for the 
$R_{nl}({r})$,
and
introduce a non-integer quantum number $\nu$,
\begin{eqnarray}
E_{3D} = \left(2 \nu + l + \frac{3}{2} \right)\hbar \omega_{ho}.
\label{eq_nu3D}
\end{eqnarray}
For $l=1$, we obtain
\begin{eqnarray}
\frac{1}{3 \sqrt{\pi^3}} 
\left[\frac{1}{2} 
\frac{\partial^3}{\partial r^3}
\left( \exp\left(-\frac{r^2}{2 a_{ho}^2}\right) r^3
\sum_{n=0}^{\infty}  \frac{
L_n^{(3/2)}(r^2/a_{ho}^2)}{n - \nu} 
 \right)
\right] _{{r} \rightarrow 0}
= \nonumber \\
 -
\frac{\hbar \omega_{ho} \; a_{ho}^5}{g_1}.
\label{eq_sum3Dc}
\end{eqnarray}
Using the identity
\begin{eqnarray}
\sum_{n=0}^{\infty} \frac{L_n^{(3/2)}\left(r^2/a_{ho}^2\right)}{n-\nu}=
\Gamma(-\nu) \;U\left(-\nu,\frac{5}{2},\frac{r^2}{a_{ho}^2}\right),
\end{eqnarray}
the infinite sum in Eq.~(\ref{eq_sum3Dc}) can be rewritten,
\begin{eqnarray}
\frac{\Gamma(-\nu) }{3\sqrt{\pi^3}} 
\left[\frac{1}{2} 
\frac{\partial^3}{\partial r^3}\left( \exp\left(-\frac{r^2}{2 a_{ho}}
\right) r^3
\;\; U\left(-\nu,\frac{5}{2},\frac{r^2}{a_{ho}^2}\right)
 \right)
\right] _{{r} \rightarrow 0}
= \nonumber \\
 -
\frac{\hbar \omega_{ho} \; a_{ho}^5 }{g_1},
\label{eq_interm}
\end{eqnarray}
where the $r \rightarrow 0$ limit 
is, as discussed above, 
due to the regularization operator of $V^{l=1}_{pseudo}(\vec{r})$ 
well behaved.
Expression~(\ref{eq_interm}) can be evaluated using the known small $r$
behavior of the hypergeometric function 
$U ( -\nu,\frac{5}{2},\frac{r^2}{a_{ho}^2})$~\cite{abraS1},
\begin{eqnarray}
\frac{1}{\pi} \Gamma(-\nu) \; 
U\left(-\nu,\frac{5}{2},\frac{r^2}{a_{ho}^2}\right) \rightarrow
- \left(\frac{r}{a_{ho}}\right)^{-3}
\frac{1}{\Gamma(-\frac{1}{2})} 
\nonumber \\
- \left( \frac{r}{a_{ho}} \right) ^{-1}
\frac{(2 \nu + 3)}{\Gamma(-\frac{1}{2})} +
\frac{
\Gamma(-\nu)}{\Gamma(-\nu-\frac{3}{2}) \Gamma(\frac{5}{2})}+
{\cal{O}}(r)
.
\label{eq_ex33}
\end{eqnarray}
If we insert expansion~(\ref{eq_ex33}) into Eq.~(\ref{eq_interm}),
evaluate the derivatives,
and take the $r \rightarrow 0$ limit,
we find
\begin{eqnarray}
-
\frac{\hbar \omega_{ho} \; a_{ho}^5 }{g_1}
= 
\frac{1 }{\sqrt{\pi}} 
\frac{\Gamma(-\nu)}{\Gamma(-\nu - \frac{3}{2})\Gamma(\frac{5}{2})}.
\end{eqnarray}
Using Eqs.~(\ref{eq_couplingg1}) and (\ref{eq_nu3D}),
we obtain our final expression for the relative eigenenergies
$E_{3D}$ for $l=1$,
\begin{eqnarray}
\frac{V_p}{a_{ho}^3} = - \frac{\Gamma(-\frac{E_{3D}}{2 \hbar \omega_{ho}}-
\frac{1}{4})}
{8 \Gamma(-\frac{E_{3D}}{2 \hbar \omega_{ho}} + \frac{5}{4})}.
\label{eq_finalpwavewave}
\end{eqnarray}

\begin{figure}[tbp]
\vspace*{.3in}
\centerline{\epsfxsize=3.25in\epsfbox{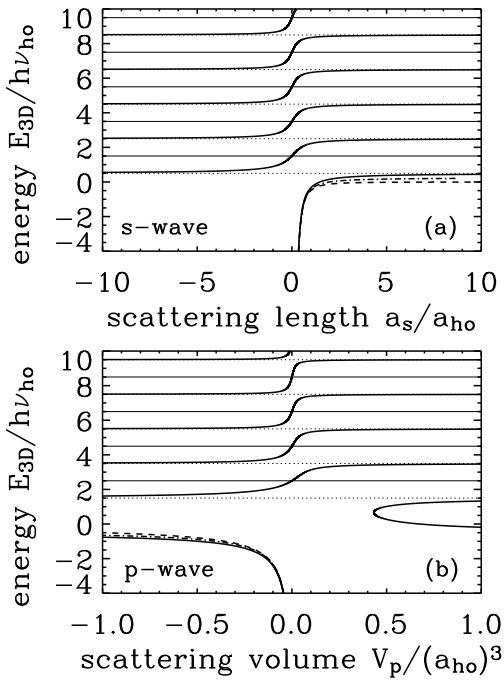}}
\vspace*{-.3in}
\caption{
Solid lines in panel (a) show
the relative $s$-wave energies $E_{3D}$  [Eq.~(\ref{eq_3Dswave})]
calculated using the pseudo-potential
$V_{pseudo}^{l=0}(\vec{r})$
as a function of the scattering length $a_{s}$.
Solid lines in panel (b) show
the relative $p$-wave energies $E_{3D}$  [Eq.~(\ref{eq_finalpwavewave})]
calculated using the pseudo-potential
$V_{pseudo}^{l=1}(\vec{r})$
as a function of the scattering volume $V_p$.
Horizontal solid lines indicate the 
harmonic oscillator eigenenergies [for $l=0$ in panel (a), and
for $l=1$ in panel (b)].
Horizontal dotted lines indicate the asymptotic eigenenergies $E_{3D}$ 
[for $a_s \rightarrow \pm \infty$ in panel (a), and 
for $V_{p} \rightarrow \pm \infty$ in panel (b)].
Dashed lines show the binding energies, 
Eq.~(\ref{eq_3Dbindswave}) in panel (a) and 
Eq.~(\ref{eq_ebindpwave}) in panel (b),
of the pseudo-potentials
$V_{pseudo}^{l=0}(\vec{r})$ and $V_{pseudo}^{l=1}(\vec{r})$, respectively,
without confinement.
Dash-dotted lines show the expansion of Eq.~(\ref{eq_3Dswave}) [panel (a)] and 
Eq.~(\ref{eq_finalpwavewave}) [panel (b)] including the next order term.
}
\label{fig3}
\end{figure}

Solid lines in Fig.~\ref{fig3}(b) show the relative
3D eigenenergies $E_{3D}$, Eq.~(\ref{eq_finalpwavewave}), 
for two spin-polarized fermions
under external harmonic confinement
interacting through the zero-range pseudo-potential
$V_{pseudo}^{l=1}(\vec{r})$ as a function of the
3D scattering volume $V_p$. 
For vanishing coupling strength $g_1$ (or equivalently, 
for $V_p=0$),
$E_{3D}$ coincides with the $l=1$ harmonic
oscillator eigenenergy. As $V_p$ increases
[decreases], $E_{3D}$ increases [decreases].

Expansion of Eq.~(\ref{eq_finalpwavewave})
for a large and negative eigenenergy (and hence negative $V_p$),
$|E_{3D}| \rightarrow \infty$, results in
\begin{eqnarray}
E_{3D} = -\frac{\hbar^2}{2 \mu (V_p)^{2/3}},
\label{eq_ebindpwave}
\end{eqnarray}
which agrees with the binding energy of $V_{pseudo}^{l=1}(\vec{r})$
without the
confinement potential.
A dashed line in 
Fig.~\ref{fig3}(b) shows this binding energy, 
while a 
dash-dotted line shows the expansion of Eq.~(\ref{eq_finalpwavewave})
including the next higher order.
Compared to
the eigenenergy of the system without confinement, Eq.~(\ref{eq_ebindpwave}),
the lowest eigenenergy given by Eq.~(\ref{eq_finalpwavewave})
is downshifted.
This downshift is somewhat counterintuitive, and contrary to the $s$-wave 
case. 

In addition to the eigenergies $E_{3D}$ of two atoms with $l=1$
under
harmonic confinement, we determine the corresponding eigenfunctions
$\psi_{3D}(\vec{r})$,
\begin{eqnarray}
\psi_{3D}(\vec{r}) \propto 
\frac{\Gamma(-\nu)}{(a_{ho})^{3/2}} 
\frac{r}{a_{ho}} \exp\left(-\frac{r^2}{2 a_{ho}^2} \right) 
U \left(-\nu,\frac{5}{2},\frac{r^2}{a_{ho}^2} \right),
\label{eq_wavefctpwave}
\end{eqnarray}
which lead to the
well-behaved boundary condition
\begin{eqnarray}
\left[ 
\frac{\frac{\partial^3}{\partial r^3} \left(\frac{1}{2} r^2 \psi_{3D}(\vec{r})
\right) }{r^2 \psi_{3D}(\vec{r}) }
\right] _{r \rightarrow 0} = -\frac{1}{V_p}.
\label{eq_bcpwave}
\end{eqnarray}
This boundary condition is an alternative
representation of the pseudo-potential $V_{pseudo}^{l=1}(\vec{r})$,
and depends on only one parameter, that is, the
scattering volume $V_p$. 
This is in contrast to earlier work~\cite{demk81,frol03},
which treated a boundary condition similar to 
Eq.~(\ref{eq_bcpwave}) but
evaluated the left hand side at a finite value of $r$, i.e., at $r=r_e$.
The boundary condition containing the finite parameter $r_e$
{\em{cannot}} be mapped to a zero-range pseudo-potential.
References~\cite{andr84,andr86,balt00} 
discuss alternative derivations and representations of 
boundary condition~(\ref{eq_bcpwave}).

\subsection{Comparison with shape-dependent 3D atom-atom potential}
\label{sec3Db} 
\begin{figure}[tbp]
\vspace*{0.3in}
\centerline{\epsfxsize=3.25in\epsfbox{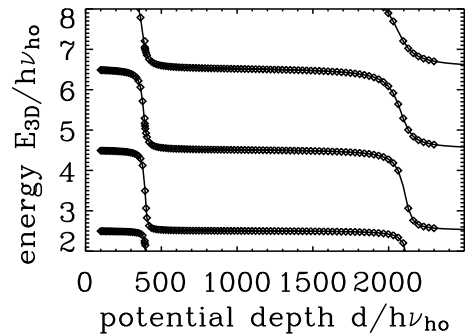}}
\vspace*{-2.1in}
\caption{
Relative 3D eigenenergies $E_{3D}$ with $l=1$ for two
spin-polarized fermions under 3D harmonic confinement as a function
of the well depth $d$.
Solid lines show the eigenenergies obtained by solving the
3D Schr\"odinger equation, Eq.~(\ref{eq_se3D3D}), 
for the Hamiltonian given in Eq.~(\ref{eq_h3D}) 
numerically
for
a short-range model potential, Eq.~(\ref{eq_1Dmorse})
with ${z}$ replaced by ${r}$ and ${z}_0$ 
replaced by ${r}_0$, for a series
of well depths $d$.
Symbols show the eigenenergies obtained for the pseudo-potential
$V_{pseudo}^{l=1}(\vec{r})$, 
taking the
energy-dependence of the 3D scattering volume $V_{p}$ into account,
$V_{p}=V_{p}(E_{sc})$ (see text). 
}
\label{fig4}
\end{figure}

To benchmark our $p$-wave
pseudo-potential treatment of two spin-polarized
3D fermions under harmonic
confinement, we solve the 3D Schr\"odinger equation,
Eq.~(\ref{eq_se3D3D}), for the Hamiltonian given by 
Eq.~(\ref{eq_h3D}) numerically for the shape-dependent Morse potential
$V_{morse}({r})$,
Eq.~(\ref{eq_1Dmorse}) with ${z}$ replaced by ${r}$ 
and ${z}_0$ replaced by ${r}_0$.
As in Sec.~\ref{sec1Db}, 
our numerical calculations are performed for 
${r}_0=11.65$a.u.,  $\alpha=0.35$a.u., 
$\omega_{ho}=10^{-9}$a.u.
($2 \pi \, \nu_{ho} = \omega_{ho}$), and 
$m=m(^{87}$Rb).
The well depth $d$
is chosen such that the 3D
Morse potential supports between zero and two $l=1$
bound states.
Solid lines in Fig.~\ref{fig4} show the resulting
3D eigenenergies $E_{3D}$ with $l=1$ 
obtained numerically
as a function of the depth $d$.

To compare the $l=1$ eigenenergies obtained numerically for the Morse
potential $V_{morse}(r)$
with those obtained for the $p$-wave pseudo-potential 
$V_{pseudo}^{l=1}(\vec{r})$, we 
follow the procedure outlined in Sec.~\ref{sec1Db}, that is,
we first determine the energy-dependent free-space
scattering volume $V_p(E_{sc})$,
$V_p(E_{sc}) = -\tan(\delta_p(k))/ k^3$,
for the
3D Morse potential (no confinement) as a function of the 
relative scattering energy 
$E_{sc}$
for various well depths $d$. 
We then solve the 
transcendental equation~(\ref{eq_finalpwavewave}) 
self-consistently
for 
$E_{3D}$. 
Diamonds in Fig.~\ref{fig4} show the resulting $l=1$
eigenenergies $E_{3D}$ for
two 3D particles under harmonic confinement
interacting through the $l=1$
energy-dependent pseudo-potential $V_{pseudo}^{l=1}(\vec{r})$ with 
$V_{p}=V_{p}(E_{sc})$.
Excellent agreement between these eigenenergies
and those 
obtained for the Morse potential (solid lines) 
is visible for all well depths $d$.
We emphasize that this 
agreement depends crucially on the usage of
{\em{energy-dependent}} 3D scattering volumes.
Figure~\ref{fig4} illustrates that the $p$-wave pseudo-potential 
$V_{pseudo}^{l=1}(\vec{r})$
describes $p$-wave scattering processes properly.

\section{Summary}
We determined the eigenspectrum for two 1D particles under
harmonic confinement interacting through a momentum-dependent 
zero-range potential. This pseudo-potential
acts only on states with odd-parity, and is hence applicable to
the scattering between two spin-polarized 1D fermions.
We showed that a basis set expansion in continuous functions
can be used to determine the eigenenergies and discontinuous
eigenfunctions of two 1D particles under harmonic confinement
interacting through the odd-parity pseudo-potential $V_{pseudo}^-(z)$.
Our divergence-free treatment
confirms the Fermi-Bose duality in 1D for two particles.

We also determined an implicit expression for the eigenenergies 
$E_{3D}$, Eq.~(\ref{eq_finalpwavewave}), and eigenfunctions
$\psi_{3D}(\vec{r})$, Eq.~(\ref{eq_wavefctpwave}), of
two 
spin-polarized 3D fermions 
under harmonic confinement interacting through a momentum-dependent
zero-range potential. 
Similar to studies of two atoms with $l=0$~\cite{ties00,blum02,bold02,borc03},
our analytical expressions might be useful in understanding the
behavior of two confined spin-aligned fermions, including 
physics near Feshbach resonances.
The $p$-wave pseudo-potential used in our study
contains derivative operators as well as a regularization operator;
the former is needed to construct a true zero-range potential
(since $l=1$ solutions go to zero as $r$ approaches zero, see above) while
the latter is needed to eliminate divergencies of 
the irregular $p$-wave solution (which diverges as $r^{-2}$).
We showed that our zero-range potential $V_{pseudo}^{l=1}(\vec{r})$
imposes a boundary condition at $r=0$,
Eq.~(\ref{eq_bcpwave}), which depends on a single atomic physics
parameter, that is,
the scattering volume $V_p$. This boundary condition is an alternative
representation of $V_{pseudo}^{l=1}(\vec{r})$.

Similarly to Fermi-Huang's regularized $s$-wave pseudo-potential, 
the $p$-wave pseudo-potential used here might find 
applications in developing 
effective many-body theories for ultracold 
spin-polarized Fermi gases. Such theories will have to carefully
investigate how to implement renormalization procedures needed
in numerical calculations.

Note added:
After submission of this paper we became aware of a related
study by Stock~{\em{et al.}}, see quant-ph/0405153, which derives 
Eq.~(\ref{eq_finalpwavewave}) starting with a
pseudo-potential expressed as the limit of a $\delta$-shell.

\acknowledgements
This work was
supported by the NSF under grant PHY-0331529.
Discussions with 
Dimitri Fedorov, 
Marvin Girardeau and Brian Granger are gratefully acknowledged.


\end{document}